\documentclass[twoside,11pt]{article}

\usepackage{jmlr2e_modified_no_mention_of_journal} 
\usepackage{times}
\usepackage{algorithm}
\usepackage{algorithmic}
\usepackage{amsfonts}
\usepackage{amsmath}
\usepackage{latexsym}
\usepackage{color}
\usepackage{graphicx}
\graphicspath{ {./images/} }
\usepackage{enumerate}
\usepackage{amstext}
\usepackage{url}
\usepackage{hyperref}

\DeclareMathOperator*{\E}{\rm E}
\DeclareMathOperator*{\argmax}{\rm argmax}

\newcommand{\ignore}[1]{}

\newenvironment{proof*}{\trivlist
\item[\hskip\labelsep{\it\proofname}{.}]}

\jmlrheading{}{2018}{}{}{}{Greenberg}
 \ShortHeadings{Calibration Scoring Rules for Practical Prediction Training}{Greenberg}
\firstpageno{1}

\begin{document}

\title{Calibration Scoring Rules for Practical Prediction Training}

\author{\name Spencer Greenberg 
      \\\addr ClearerThinking.org
       \\\email spencer@clearerthinking.org
        }
       
\editor{}

\maketitle

\begin{abstract}
In situations where forecasters are scored on the quality of their probabilistic predictions, it is standard to use `proper' scoring rules to perform such scoring. These rules are desirable because they give forecasters no incentive to lie about their probabilistic beliefs. However, in the real world context of creating a training program designed to help people improve calibration through prediction practice, there are a variety of desirable traits for scoring rules that go beyond properness. These potentially may have a substantial impact on the user experience, usability of the program, or efficiency of learning. The space of proper scoring rules is too broad, in the sense that most proper scoring rules lack these other desirable properties. On the other hand, the space of proper scoring rules is potentially also too narrow, in the sense that we may sometimes choose to give up properness when it conflicts with other properties that are even more desirable from the point of view of usability and effective training. We introduce a class of scoring rules that we call `Practical' scoring rules, designed to be intuitive to users in the context of `right' vs. `wrong' probabilistic predictions. We also introduce two specific scoring rules for prediction intervals, the `Distance' and `Order of magnitude' rules. These rules are designed to satisfy a variety of properties that, based on user testing, we believe are desirable for applied calibration training.
\end{abstract}


\section*{Acknowledgements}
Thanks to \href{https://yngve.hoiseth.net}{Yngve Hoiseth} for making the charts (using \href{https://github.com/yhoiseth/python-prediction-scorer}{Python Prediction Scorer}).

\section{Introduction}
We consider situations where a forecaster is making probabilistic predictions, and the goal is to assign points to each prediction that will be shown to the forecaster, in order to help the forecaster improve their future predictions. We consider this question in the real world context of creating an online calibration training program (the `Calibrate Your Judgment' app, in collaboration with the Open Philanthropy Project). In this digital program, users practice making probabilistic predictions on thousands of questions across a variety of domains (e.g. trivia, simple math problems, politics, correlations visually determined from a scatter plot, etc.) and receive a score immediately after each prediction indicating the quality of that prediction. Users can also view their average performance over time, as well as their `calibration curve', showing what percent of the time their predictions are correct (y) plotted against the chance that they assigned to their predictions being correct (x). In this context of a real world training system, where the goals are to maximize the speed at which users learn, as well creating a comprehensible and enjoyable user experience, there are numerous considerations that suggest significant modifications to the standard rules for scoring probabilistic predictions.
\section{Scoring probabilities}
It is easy to give continuous performance feedback to a forecaster who merely predicts whether an event will, or will not occur. One can simply tell the forecaster whether each prediction was correct or incorrect.
\\
\\
However, if the forecaster assigns probabilities to events, or provides a range for numerical estimates (i.e. a prediction interval) we would like to be able to incorporate these probabilities (or this interval) in the points we assign to each prediction. In particular, we need our feedback to give the forecaster a meaningful assessment of how good their prediction was as soon as each forecast resolves (i.e. in real time if the answer is already known, and as soon as the answer is known in other cases). For instance, if two forecasters have the same accuracy, but one is correct more often when confident and less often when unconfident, whereas the other is correct equally often when confident as when unconfident, the former is a better predictor. Hence, the confidence level of the forecaster should be taken into account when assigning points to each prediction. But how should these points be assigned?
\\
\\
There is an existing mathematical theory of how to assign points to probabilistic predictions, namely the theory of `proper scoring rules'. These scoring rules are appealing because they give no incentive for the forecaster to lie about their probabilistic beliefs. However, in a real world system intended to actually train forecasters, this is one desirable property among others (albeit an important one). 
\\
\\
\section{Desirable properties of scoring rules}

Since our goal is a pragmatic rather than theoretical one, namely helping the forecaster learn to be as good a predictor as possible in as little time as possible (and with a pleasant user experience in the process), our criteria for scoring rules do not only rely solely on formal mathematical properties such as `properness'. Learning is a psychological process, and so our criteria necessarily involve assumptions about the psychology of users. To perhaps a surprising degree, the typical scoring rules used in probabilistic forecasting based solely on desirable mathematical properties do not, in our experience, produce performance feedback that has the intuitive properties a forecaster would expect.

Based on our experience building and testing an online calibration training program, we view it as desirable for a scoring rule to have all of the following properties:

 \begin{enumerate}

\item \textbf{Good prediction performance should be rewarded}. That is, the forecaster should receive more points for better predictions. In practice that means that when a forecaster has made a correct prediction, they should receive more points the more confident they are, and when they have made an incorrect prediction they should receive fewer points (or lose more points) the more confident they are. In the context of a prediction interval given by the forecaster to predict a numerical value, when the true answer is close to the center of the prediction interval the forecaster should receive more points the narrower the interval is. Whereas when the true answer is far from the middle of the prediction interval the forecaster should receive fewer points (or lose more points) the narrower the interval is.

\item \textbf{Forecasters should be able to quickly `process' the points they receive}. If the user has to carefully think about the points they were assigned merely to understand whether they made a good or bad prediction, we anticipate this may slow down learning in two ways. First, it will add time to the training process, meaning that the forecaster can complete fewer predictions in a given amount of training time. Second, in the context where real-time feedback is being given (because the answers to the forecasts are already known), if the player's brain can process the points as a reward or punishment within seconds of each decision being made, it may accelerate learning via the mechanism of operant conditioning. Numerous studies in the animal learning literature show that learning is faster when feedback occurs more quickly (and that if feedback is too slow, learning may not occur at all). If a dog is rewarded one hour after a desirable behavior, it may never learn to associate the behavior with the reward. Even a five minute delay may be too long to create the desired low level association between behavior and reward. But if mere seconds elapse between the behavior and the reward, the dog will learn quickly to perform the behavior, because its brain will be able to reliably predict the reward as a consequence of engaging in the behavior. Hence, we speculate that if the forecaster's brain can immediately process how good each of their predictions was, this may accelerate the learning process by engaging low level learning processes that can't operate with long delays. Of course, humans are entirely capable of learning prediction tasks that have long feedback delays, but we believe that learning efficiency will be maximized when the delay between action and reward (or punishment) is eliminated.

\item \textbf{The scoring should feel intuitive to forecasters}. When user tested various versions of our scoring rules, we found it is easy to confuse users if the point scores assigned don't conform to certain expected intuitive properties. To avoid this confusion, hopefully making the learning process both more enjoyable and more efficient, we put considerable effort into designing scoring rules that have the properties that we believe users will often assume the scoring system will satisfy.

\item \textbf{Overall performance should not be too influenced by a small subset of predictions.} Some point scoring rules can produce arbitrarily large changes in a forecaster's points. Outliers like these are undesirable for at least two reasons. First, they mean that a forecaster may not be able to `recover' from a very bad prediction, or may feel low motivation to keep playing if one prediction produces so many points that subsequent predictions appear to produce trivially small point changes in comparison. Second, if we are comparing forecasters to each other (e.g. by their average points per forecast) we want their scores to be driven by a large number of predictions, not by one or two massive outliers (e.g. their one or two worst predictions). When large point outliers can occur it may take many more predictions to converge to an accurate assessment of any given forecaster's skill.

\item \textbf{Honest belief reporting should be incentivized.} Since the purpose of training is to teach forecasters to make accurate predictions, it is undesirable if the point scoring system is `gameable', meaning that it incentivizes forecasters to lie about their actual probabilistic beliefs. The theory of proper scoring rules provides us with a solution to this problem, though as we will see, these scoring rules do not necessarily make it easy to satisfy our other desirable criteria for point scoring.
\end{enumerate}

\section{Properties of an intuitive scoring rule}
But what does the `intuitiveness' of a scoring rule mean? In our testing of scoring rule variants, we identified a number of properties that may be desirable from the perspective of making a scoring real feel more intuitive to users, namely:
\\
\\
(1) \textbf{Higher is better} - A higher score should correspond to making a `better' prediction. That way, your score can be thought of as indicating how good your prediction was. This corresponds to the very common practice of awarding points in video games, where the better you perform the more points you get. While a training system could be the case that the scoring rule assigns `error' instead of `points' (as is the case with some proper scoring rules), we think most users will anticipate higher numbers as corresponding to better performance, rather than the reverse.
\\
\\
(2) \textbf{Upper bound} - There is a finite upper bound on the number of points you can get on a single prediction, and this maximum should occur when you make the best possible prediction (i.e. you are correct and as confident as possible). As previously discussed, an unbounded number of points on a single prediction is problematic because it means that a single prediction can dominate a player's total score (e.g. if there was no bound, then one amazing prediction could potentially give so many points as to make all your other predictions irrelevant). What's more, by having the `best possible' prediction always give a known and fixed number of points, that makes performance across different types of prediction tasks or different prediction categories more comparable. Yet another reason that an upper bound is desirable is that it pushes the forecaster to spreading out their forecasting effort across many predictions, rather than focussing their mental energy and concentration on forecasts where they think they could potentially earn an enormous number of points.
\\
\\
(3) \textbf{Lower bound} - There is finite lower bound on the number of points you can get in a single prediction (or lose, in which case we're talking about the most negative points possible per prediction). This minimum should occur when you make the worst possible prediction (i.e. you are incorrect and as confident as possible). This lower bound prevents one really bad prediction from hurting your score so much that all your other playing becomes irrelevant. On the other hand, we do not view it as necessarily obvious that one should have the lower bound be the same value across prediction types. The reason being that some prediction types can have answers that might be considered more wrong than is possible in other prediction categories. For instance, if you are trying to determine which of two mutually exclusive options will occur, the most wrong you can be is to pick the wrong option of the two options possible, and be totally confident in that mistaken choice. If you are 100\% confident in your wrong answer, there is a sense in which this prediction is maximally wrong. But one might argue that it is intuitively still not as wrong as you can be in the prediction interval case, where you can predict for example that a numerical value falls in the range 1 billion to 1.000000001 billion with 90\% confidence when the real value is just 10, which perhaps may be considered more deeply wrong than being totally confident and wrong in the true/false case.
\\
\\
(4) \textbf{Positive is good, negative is bad} - A positive number of points should only occur if you get the prediction `correct', and a negative number of points should only occur if you get the prediction `incorrect'. In our experience testing different point systems it can be quite confusing for users to gain points for an incorrect prediction or lose points for a correct prediction. Similarly, in the case of prediction intervals, our testing suggested that it can be confusing to forecasters if you do not receive positive points when the true value falls within your prediction interval. More generally, we think it is natural for forecasters to think of it being good to gain points, and bad to lose points, with zero points corresponding to a neutral outcome. Hence, we ideally want the sign of the points we give to correspond to this intuition. Doing so also clearly communicates to users whether they did a `good' job on their prediction overall or not, which may speed up how quickly they can process the reward or punishment of the points they are given (a positive number is a reward, a negative number is a punishment, and the magnitude is the size of the reward or punishment), thus potentially accelerating learning. 
\\
\\
(5) \textbf{Confidence is good when correct, bad when incorrect} - When you are `correct', you should gain more points for a more confident prediction, and when you are `incorrect' you should lose more points for a more confident prediction. This property is important because it teaches the critical skill of learning to feel confident when you are likely to be right, and to have low confidence when you are likely to be wrong. It is also a property shared by the common proper scoring rules such as the quadratic, log, and (negative) brier score rules.
\\
\\
(6) \textbf{Total uncertainty should give zero points} - If your prediction has the maximum amount of uncertainty possible, then we think it is intuitive to receive 0 points, since in this case the forecaster really has no forecast at all, and so is neither right nor wrong. Essentially, it is a case of a forecaster indicating that they have no information with which to make a forecast (e.g. indicating 50\% confidence in a binary `yes' vs. `no' prediction). Note that this property has a side effect, namely that it provides a convenient mechanism by which players can skip questions that they don't want to make a prediction about (by indicating the maximum uncertainty possible). In the case of prediction interval forecasts, our testing suggested to us that it is natural to think of zero points corresponding to the true value falling just barely outside of the prediction interval. So in other words, if the prediction interval were [10, 100] and the true value were just below 10, users might expect to receive approximately zero points.
\\
\\
(7) \textbf{Continuity} - An arbitrarily small change in your prediction should create only a small change in your score (i.e. the scoring function should be continuous with respect to your predicted probability or prediction interval). We think it would be surprising to forecasters if they were able to get a substantial increase in points by adjusting their estimated probability from, say, $0.90$ to $0.901$. Since small changes in probabilities or prediction intervals correspond to small changes in beliefs, it does not intuitively seem that such small changes should cause large changes in points awarded. However, while we think this is an intuitive property that users will anticipate, we also think that a reasonable violation of this property is to round the scores to the nearest integer for display purposes (i.e. users don't want or need to see their points to a large number of significant digits). 
\\
\\

\section{Mathematical formulation}

For now on, we will use the phrase `scoring rule' in a formal sense, to mean any function that assigns points to a prediction intending to indicate the quality of that prediction. We assume that points are represented by real numbers, and that the options that the forecaster chooses among are mutually exclusive and collectively exhaustive (i.e. one and only one such option will be the actual answer). Scoring rules will take into account both the (probabilistic) prediction of a forecaster, and what the correct answer for that forecast turned out to be.
\\
\\
In the case where the prediction is about which of a finite number of mutually exclusive options is correct (or will turn out to be correct if the event has not yet occurred), we write our scoring rule:
\begin{equation}
S(p, e)
\end{equation}
where $p = (p_1, p_2, \ldots, p_n)$ is a vector of the probabilities that the forecaster gave for the mutually exclusive events, with probability $p_i$ being assigned to outcome $i$. Note that $\sum_{i=1}^{n} p_i = 1$ since the outcomes are both mutually exclusive and collectively exhaustive (so exactly one of the possible options will turn out to be correct). Furthermore, here $e = (e_1, e_2, \ldots, e_n)$ is a vector indicating which of the mutually exclusive options turned out to be the correct one, with $e_c = 1$ for the correct option c, and $e_j = 0$ for all $j \ne c$.
\\\\
\subsection{Prediction interval forecasts}
If the prediction problem is such that the forecaster provides a real valued range [L,U] that he or she believes some numerical outcome will be contained within with a fixed probability $\beta$, i.e. a prediction interval forecast, then we write our scoring rule: 

\begin{equation}
S(x, L, U)
\end{equation}
where x is the actual (true) numerical value that the variable the forecaster is predicting turns out to be.
\\\\

\subsection{Proper Scoring Rules}

In an ideal world, we would want our scoring rule to be `proper'. A proper scoring rule is any scoring rule where a forecaster's average point score is maximized (assuming higher scores are better) when the forecaster predicts using the probabilities $p$ that they actually believe are correct. Or, in the case of a prediction interval forecast, when they provide their true range [L, U] that they believe the true value actually has a $\beta$ probability of falling within. In other words, proper scoring rules provide no incentive to the forecaster to lie about their probabilistic beliefs. A scoring that is not proper has the drawback that a forecaster may be incentivized to predict a probabilities other than the ones they actually believe. A `strictly' proper scoring rule is even better than a proper scoring rule, since it has the additional constraint that the \textit{only} way for the forecaster to maximize their score is to give their `true' probabilities or prediction interval (i.e. their honest predictions are a unique maximizer of the score's expected value).
\\
\\
If, for a probabilistic prediction, we let $\bar p$ denote the vector of probabilities that the forecaster believes are accurate specifications of the likelihood of the mutually exclusive options occurring (in contrast to $p$, which is the vector probabilities \textit{submitted} by the forecaster for scoring by our scoring rule, which may or may not equal $\bar p$), a proper scoring rule is one satisfying:

\begin{equation}
\bar p = \argmax_{p} \E_{\bar p} [ S(p, e) ]
\end{equation}
where $\E_{\bar p} [\ldots]$ is the expected value (over $e$) with respect to the probability distribution $\bar p$, meaning that $e_i$ has a probability $\bar p_i$ of being a 1 (with all other entries of $e$ then necessarily being 0 due to the options being mutually exclusive). In words, this equation means that the forecaster gets the highest average score if they submit as their probabilities the ones they believe are actually true accounts of the chance that those mutually exclusive options occur.

\section{Popular proper scoring rules}
\label{sec:preliminaries_ubound}

While there are an infinite number of proper scoring rules (and a `large' type of infinity, at that), there are only a few that are popular in practice. Here we consider some of the most popular strictly proper scoring rules (in their typical formulations). Furthermore, we discuss some of their properties that are potentially problematic when attempting to use these popular formations for real world prediction training.

\subsection{Quadratic scoring rule}

$S_{quad}(p, e) = \sum_{i=1}^{n} p_i  (2  e_i  - p_i)$
\\
\begin{figure}[h]
\caption{Quadratic score for probabilities $\{0.00, 0.01, \ldots, 1.00\}$}
\includegraphics[width=0.7\textwidth]{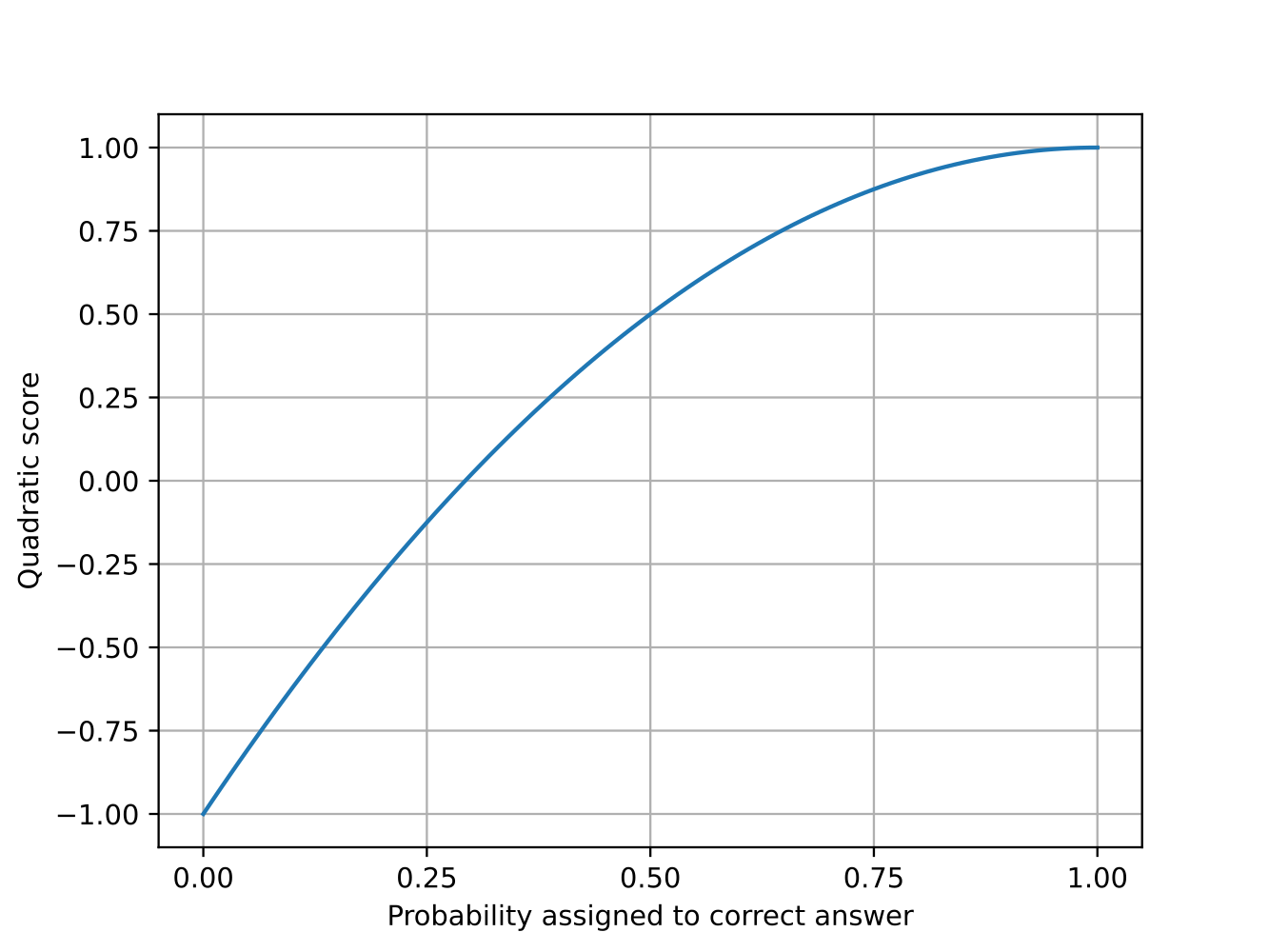}
\centering
\label{fig:quadratic}
\end{figure}
\\
Figure \ref{fig:quadratic} shows the possible values for the quadratic scoring rule. It has some desirable properties, namely that:

 \begin{itemize}
 
 \item It is strictly proper.
 
 \item There will not be outliers in the number of points assigned.
 
 \item It is simple to calculate.
 
 \item The points assigned are in the range of -1 to 1, a range that seems at least reasonably intuitive (as opposed to, say, a range of 0.3 to 0.9 which would presumably be quite unintuitive to most forecasters).
 
 \item The scoring rule can be derived from a set of four reasonably intuitive axioms \citep{selten1998axiomatic}, namely that the function is a strictly proper scoring rule, that its scores do not change if you re-order the options, that it is indifferent to adding new options that have no possibility of occuring, and that if you have two possible theories, the average score of the first theory conditioned on the truth of the second theory is the same as the average score of the second theory conditioned on the truth of the first theory (i.e. a specific form of symmetry).
 
 \end{itemize}

On the other hand, it has other properties that we view as undesirable for use in a pragmatic training system, specifically:

 \begin{itemize}
 
 \item In the true/false case (where there are only two possible options for what can be true), our experience is that users intuitively expect to receive positive points when they assign more than a 50\% chance to the correct option, and negative points when they assign more than a 50\% chance to the incorrect option. However the quadratic score, in its standard form, has the confusing property of yielding positive points when the forecaster assigns at least a probability of $1 - \frac{\sqrt 2}{2} \approx 0.29 $ to the correct answer.
 
  \item In the case where a forecaster guesses purely at random (e.g. assigning a uniform distribution over the possible outcomes), our experience is that users intuitively expect to receive no points on average (since they are in a certain sense not making a prediction at all in this case). However, the quadratic scoring rule in this case assigns a score of  $\frac{1}{n}$, where n is the number of options. 
  
  \item The quadratic scoring rule lacks certain desirable properties of the logarithmic scoring rule, as will be discussed below.
 
 \end{itemize}

\subsection{Brier scoring rule}

$S_{brier}(p, e) = \sum_{i=1}^{n} (e_i  - p_i)^2$
\\
\begin{figure}[h]
\caption{Brier score for probabilities $\{0.00, 0.01, \ldots, 1.00\}$}
\includegraphics[width=0.7\textwidth]{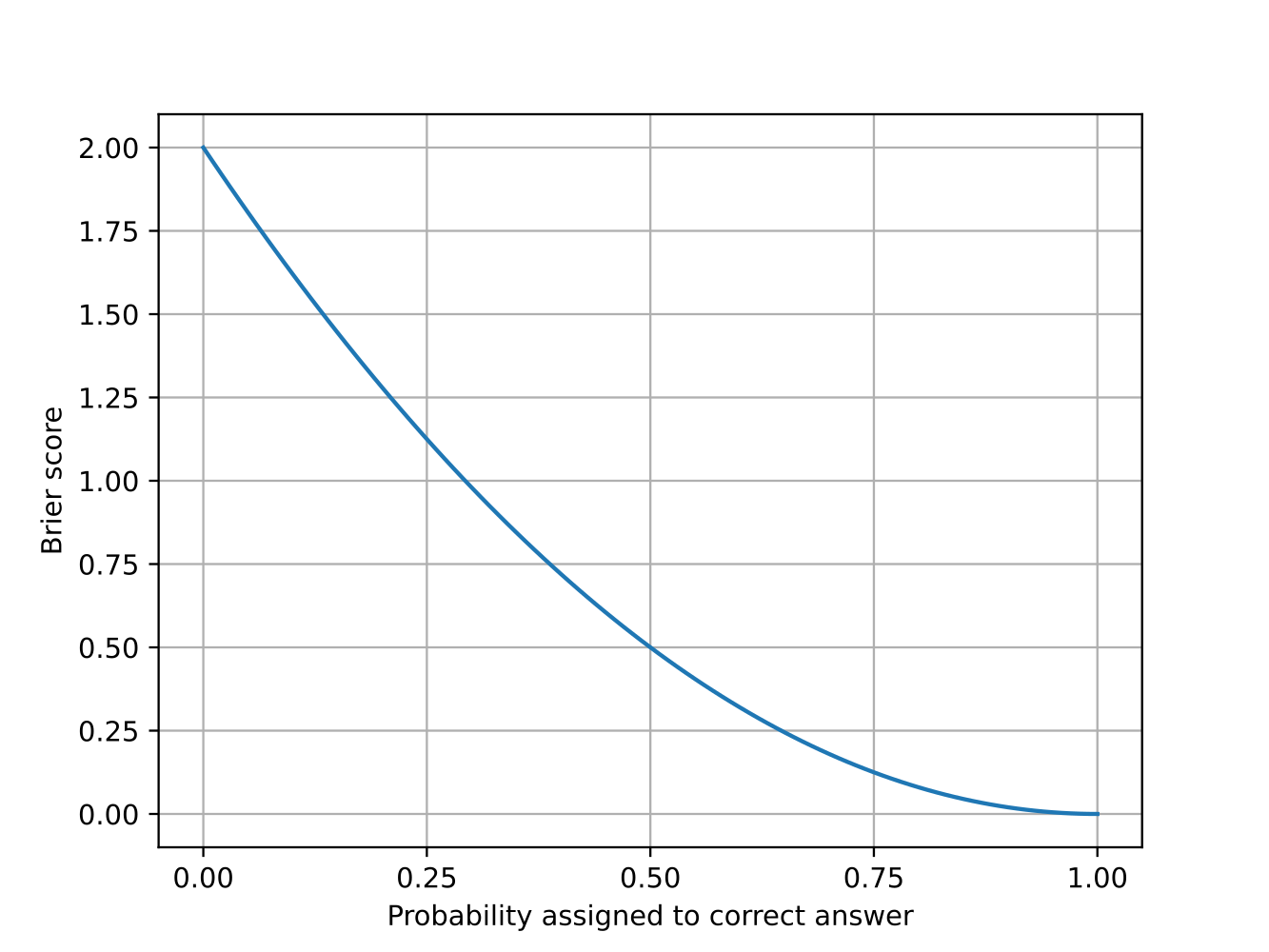}
\centering
\label{fig:brier}
\end{figure}
\\
The Brier Score — shown in figure \ref{fig:brier} — is a linear transformation of the Quadratic Scoring Rule, and hence very similar to it. However, in its standard formulation it is perhaps even less intuitive than the quadratic scoring rule, since the best possible point score you can achieve with it is 0, with positive scores indicating worse performance. Because of its similarity to the quadratic scoring rule, we will not discuss it further.
\\
\\

\subsection{Logarithmic scoring rule}

$S_{log}(p, e) = - \sum_{i=1}^{n} e_i  \log(p_i)$
\\
\begin{figure}[h]
\caption{Logarithmic score for probabilities $\{0.01, 0.02, \ldots, 1.00\}$}
\includegraphics[width=0.7\textwidth]{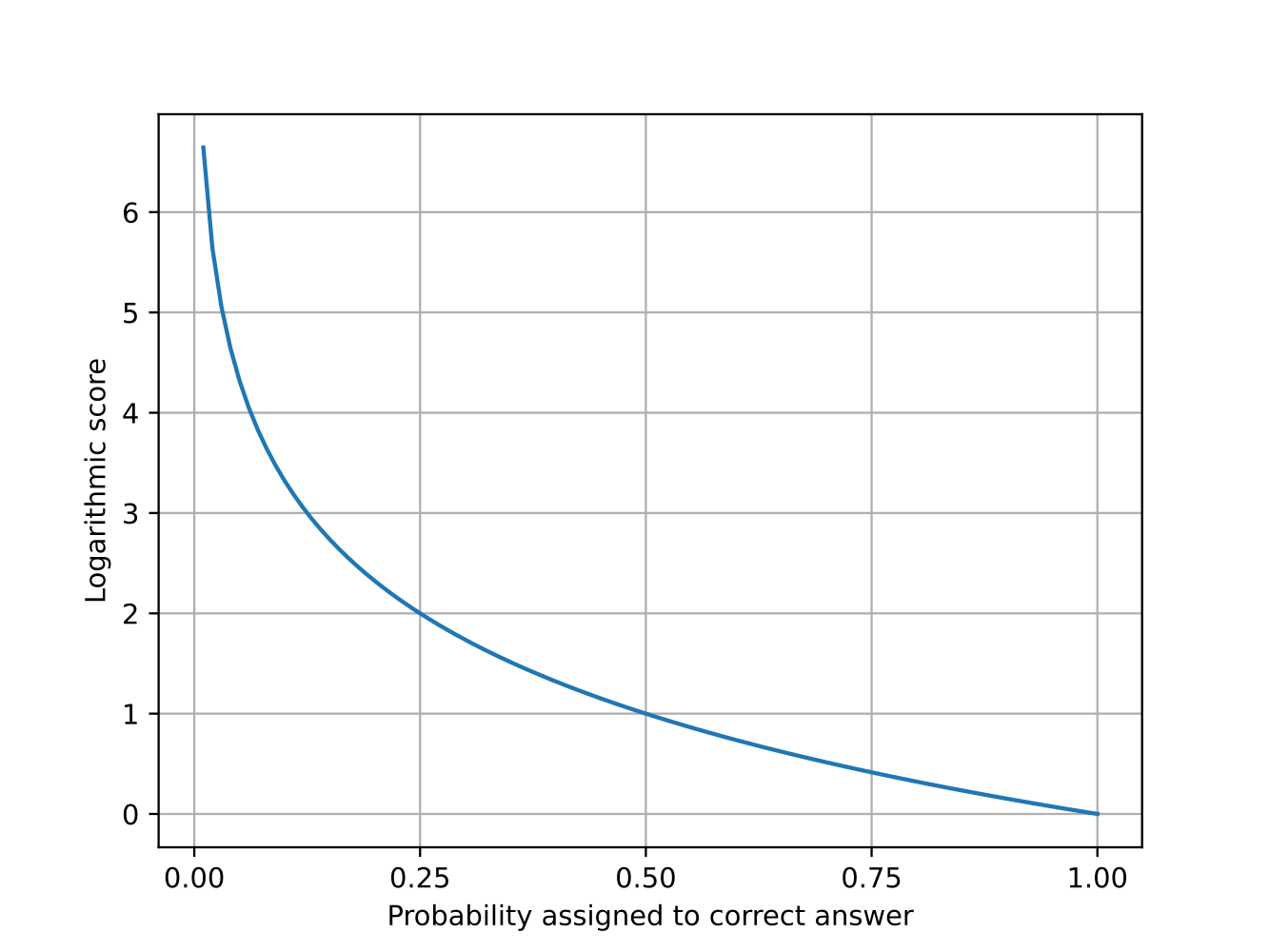}
\centering
\label{fig:logarithmic}
\end{figure}
\\

The log score — shown in figure \ref{fig:logarithmic} — has some desirable properties, namely that:

 \begin{itemize}

 \item It is strictly proper.
 
 \item It has a very nice, intuitive interpretation. The log score (when summed over multiple predictions) is simply a measure of how likely \textit{all} of those outcomes that actually occurred were, according to the probabilities that the forecaster assigned to each outcome, assuming the outcomes are all statistically independent. Why? Well, suppose that a guesser assigns probability $p_{c}^{j}$ to the correct outcome c, for their jth prediction. Then if the events are statistically independent, the probability this guesser assigns to \textit{all} the outcomes occurring that actually did occur, is given by $prob_{all} = p_{c}^{1}  * p_{c}^{2}  * p_{c}^{3} * \hdots$. Since the log function is strictly increasing, we can apply it to this probability of all the events without changing the rank order of different guessers (in other words, the score each forecaster gets will change, but if one had a bigger score before, they will still have a bigger score after we take the log of everyone's scores). Applying the rule of logarithms that $\log(a) + \log(b) = \log(a*b)$, this yields the formula: 
\\
\\
 $\log(prob_{all}) = \log(p_{c}^{1} * p_{c}^{2} * p_{c}^{3} * ...) = \log(p_{c}^{1}) + \log(p_{c}^{2}) + \log(p_{c}^{3}) + \ldots$
 \\
 \\
 This formula is precisely equal to the log score of a forecaster when we sum their points for all of the predictions! Hence, it really is just a measure of how probable the forecaster said all of the outcomes that actually happened were (assuming they are statistically independent events).
 
 \item Related to the previous property, the log score has an interesting property that it doesn't matter if you make predictions for event A then event B or for the joint even consisting of both event A and event B happening. Either way you get the same log score in the end. So if forecaster 1 predicts probability $p_a$ for the correct outcome of event A and $p_b$ for the correct outcome of event B (assuming they are statistically independent events), and forecaster 2 predicts the single probability $p_a*p_b$ for (what we now call) a single event corresponding to the correct outcome of both A and B, the two forecasters end up with same log score!
 
 \item The log score has an interpretation in terms of \textit{bits of information}, since (as used in the theory of Shannon entropy) the function $-\log(p)$ tells us how much information we learn from an event occurring (if that event occurs with probability $p$).
 
 \item In a probabilistic betting context, there is a way of interpreting the log score as the utility achieved by a bettor who has a logarithmic utility function. This is a fairly common model for utility functions that may reasonably accurately model certain real world relationships that are believed to have diminishing marginal returns (e.g. life satisfaction as a function of income).

\item In cases where probabilities are being assigned to more than two possible options, the log score (up to a linear transformation) is the only strictly proper rule that satisfies `locality' (i.e. the score depends only on the probability you assign to the single outcome that ended up happening, not on the probabilities for outcomes that didn't happen).
 
 \end{itemize}

On the other hand, the log score has other properties that we think are undesirable for use in a pragmatic training system, specifically:

 \begin{itemize}
 
 \item In its standard formulations the best possible number of points you can get is 0 (with larger numbers of points being worse), or, if the sign of the formula is reversed as it sometimes is, the formula always gives non-positive numbers of points. We think neither of these cases is likely to be intuitive to users in the real world.

 \item It will produce an infinite score if a forecaster is totally certain but guesses wrong. Even if you disallow this scenario (by banning predicting a probability of precisely 1), it can still produce arbitrarily large numbers of points just for making one really bad guess, which a guesser's total score may never be able to recover from. This is especially undesirable in a training context, where we expect people to sometimes predict really badly when they are fist starting, and we want them to be able to recover (without spending ages) from these bad predictions.
 
 \item Related to the point above, we speculate that negative outliers (such as produced by the log score) could cause people to become too afraid of  giving high confidence predictions during training. For instance, after a few times when they lose enormous numbers points for being highly confident and wrong, real players may overreact (much like a person might be overly afraid of dogs once they've been bitten once). This might lead to excessively conservative play. Psychologically, a large loss might loom greater in the forecaster's mind than an equivalently sized gain.
 
 \end{itemize}

\subsection{Other options}

Plenty of other proper scoring rules exist in the literature, such as the `spherical scoring rule', and families of rules that let you generate scoring rules by varying parameters. For instance, the `beta', `pseudospherical' or `power' families. The idea is: you specify a couple of real numbers as inputs and these families give you a single proper scoring rule as the output. Rules of thumb exist in some cases about what, intuitively, these numbers you input mean. Usually these families generalize the log score and brier score, in the sense that there are input values that cause the families to output the log score and brier score as special cases. In the simple case where there are only two possible options for the forecaster to choose between (and assign probabilities to), usually the input parameters relate to how you treat false positive vs. false negatives (i.e. the symmetry with respect to the two outcomes).
\\
\\
It is also possible to construct proper scoring rules by combining existing ones. For instance, you can multiply a scoring rule by any fixed constant, and add any other fixed constant as well (i.e. you can do a linear transformation of the scoring rule's output), and it will remain a proper scoring rule. You can also add two proper scoring rules together and still have a proper scoring rule (assuming that they are both rules where "higher is better" or they are both rules where "lower is better").
\\
\\
We can even construct a proper scoring rule from any function that satisfies a few simple properties. To illustrate this point, suppose we are in a true/false prediction setting where there are only two outcomes, and we have any function $f(x)$ of our choosing that is differentiable and convex. Then we can create a proper scoring rule $S(p_0, e_0)$ where $p_0$ is the probability assigned to that outcome being true (as opposed to false), and the scalar $e_0$ satisfies $e_0=1$ if the outcome is true, and $e_0=0$ if it is false. The proper scoring rule can then be constructed using the following simple formulas:
\\
\\
$S(p_0,1) = f(p_0) + (1­-p_0) f'(p_0)$
\\
\\
$S(p_0, 0) = f(p_0) ­ p_0 f'(p_0)$
\\
\\
where $f'(x)$ is the derivative of $f(x)$ with respect to $x$. In this sense, there are `as many' proper scoring rules as there are convex, differentiable functions.

\section{Prediction intervals}

We have discussed common options for scoring rules for cases where there are a finite number of possible outcomes. But what scoring rules can be used for prediction intervals? From \cite{gneiting} we have:
\begin{theorem}
If each $s_i$ is nondecreasing for $i=1,\ldots,k$ and $u$ is an arbitrary function, the scoring rule:
\\
\\
$S(x, r_1,\ldots, r_k) = \sum_{i=1}^{k} (\alpha_i s_i(r) + (s_i(x) - s_i(r_i)) \hspace{5pt} 1_{x \le r_i}) + u(x)$
\\
\\
is proper for predicting the quantiles at levels $\alpha_1, \ldots, \alpha_k \in (0,1)$.
\end{theorem}
The authors mention that they do not know whether this formula provides `the general form of proper scoring rules for quantiles'. If it does, that is a shame, because this formula is more limited than would be desirable in real world training contexts, as we will explain.
\\
\\
In the special case of prediction intervals (one of particular interest in forecasting training), where the forecaster gives a range $[L,U]$ that they predict will contain the true value with probability $\beta$, we can reduce the above formula to:
\begin{equation}
S(x, L, U) = u(x) - \frac{1-\beta}{2} (s_2(U) - s_1(L)) - 
\begin{cases} 
      s_1(L) - s_1(x) & x<L \\
     0 & L \le x \le U \\
      s_2(x) - s_2(U) & x > U 
\end{cases}
\end{equation}
where $u(x)$ is any function, and $s_1(x)$ and $s_2(x)$ are non-decreasing functions.
\\
\\
How can we understand what this formula is doing? Well, in the case where the guess $x$ is inside the prediction interval $[L,U]$, the point score is just:
\begin{equation}
u(x) - \frac{1-\beta}{2} (s_2(U) - s_1(L))
\end{equation}
On the other hand, when $x$ is greater than $U$ (the upper bound of the prediction interval) the score is:
\begin{equation}
u(x) -  \frac{1-\beta}{2} (s_2(U) - s_1(L)) + ( s_2(x) - s_2(U)).
\end{equation}
One way to understand these formulas is to think of $s_2(a) - s_1(b)$ as a way to measure how `wide' the interval [a,b] is (as measured by some choice of non-decreasing functions $s_2$ and $s_1$). And $u(x)$ is some function that gives extra points that depend only on what the outcome happens to be (not the forecasted probabilities), which might for instance represent compensating the user extra points when the true answer is especially large (e.g. because larger values might be harder to predict a priori). 
\\
\\
From this perspective, when $x$ lies in the prediction interval, the number of points is simply proportional to the `width' of the interval [L, U], plus however many extra points $u(x)$ assigns based on whatever the true answer is. On the other hand, when the true value $x$ lies beyond the prediction interval, the points assigned depend both on the width of [L,U] and the width of [U, x] (i.e. how far x is from the edge of the prediction interval).
\\
\\
To see why this formula limits us greatly in what our scoring rules can do, observe that when $x$ lies in the prediction interval, if we fix L and look at the point loss as U increases, the point loss increases proportionally to $s_2(U)$. Similarly, in the case where $x$ lies beyond the prediction interval, if we fix L and look at the point loss as x increases, the point loss then increases proportionally to $s_2(x)$, that is, the same function, but this time applied to x. In other words, the way the point loss increases with U is just (up to a constant multiplier) identical to how it increases with increasing x! Therefore, the choice for how we treat the case when x is in $[L,U]$ determines to a great extent how points must work when x is outside of that range.

\subsection{Linear scoring for prediction intervals}
In the special (but intuitively natural) case where we choose $s_1(a) = \frac{1}{c}$, $s_2(a) = \frac{1}{c}$, and $u(x)=d$, for real constants $c$ and $d$, we get the linear prediction interval scoring rule:

\begin{equation}
 \label{eqn:linear_prediction_interval}
S(x, L, U) = d -  \left (\frac{1-\beta}{2} \left ( \frac{U - L}{c} \right) + 
\begin{cases} 
      \frac{L - x}{c} & x<L \\
     0 & L \le x \le U \\
      \frac{x - U}{c} & x > U 
\end{cases} \right ).
\end{equation}
In words, this formula assigns $d$ points by default, and then it reduces this number of points proportional to the size of the prediction interval, that is by $U-L$. Furthermore, if the true answer falls outside of the prediction interval, it also applies a penalty that is proportional to how far the true value is from the nearest edge of the prediction interval. The constant $c$ can be thought of as a scaling parameter that determines the units of the score (e.g. feet vs. meters). Hence, $\frac{U - L}{c}$ is the width of the forecaster's prediction interval measured in the units of our choice, and $\frac{L - x}{c}$ and $\frac{x - U}{c}$ measure how far the true answer lies from the prediction interval in the case where it does not fall inside the interval (measured in the same units).\\
\\
This formula has various desirable properties, namely:

 \begin{itemize}
 
 \item It is a strictly proper scoring rule.
 
 \item It represents (up to linear transformation) what is arguably the simplest special case of the general formula above for generating scoring rules for prediction intervals.
 
 \item It has a reasonably intuitive interpretation. You always get a default number of points (i.e. d), which is then penalized proportionally to the size of your prediction interval (i.e. $\frac{1-\beta}{2} (U-L)$) and, in the case where the true value falls outside the prediction interval, there is another penalty proportional to the distance from the true value to the prediction interval (i.e. either $L-x$ or $x-U$ depending on whether the true value was below or above the boundaries of the prediction interval).

\end{itemize}
 
 \subsection{Log scoring for prediction intervals}

Given the various benefits of log scoring rules in the mutually exclusive category prediction setting discussed previously, it is natural to consider the case where we choose $s_1(a) = \frac{1}{c} \log(a)$, $s_2(a) = \frac{1}{c} \log(a)$, and $u(x)=d$, for real constants $c$ and $d$. Then, we get the log prediction interval scoring rule:

\begin{equation}
 \label{eqn:log_prediction_interval}
S(x, L, U) = d -  \left ( \frac{1-\beta}{2} \frac{ \log \left (\frac{U}{L} \right) } {c} + 
\begin{cases} 
     \frac{\log(\frac{L}{x})}{c} & x<L \\
     0 & L \le x \le U \\
      \frac{\log(\frac{x}{U})}{c} & x > U 
\end{cases} \right )
\end{equation}

as discussed, for instance, in \cite{garrabrant}. 
\\
\\
Since we can think of the $\log$ function as measuring (up to rescaling by a constant) the number of order of magnitudes of its input, we can consider this formula as having a penalty based on the number of order of magnitude spanned by the prediction interval, namely $\frac{ \log \left (\frac{U}{L} \right) } {c}$, where $c$ effectively determines the base of the logarithm (i.e. what number system we're measuring order of magnitudes in). Likewise, $\frac{\log(\frac{L}{x})}{c}$ and $\frac{\log(\frac{x}{U})}{c}$ measure how many orders of magnitude away the true value was from the prediction interval (in cases where the true value isn't inside the interval).
\\
\\
This prediction interval scoring rule has various desirable properties, namely:

 \begin{itemize}
 
 \item It is a strictly proper scoring rule.
 
 \item It is unit invariant. So if the question a forecaster is asked to make a prediction about changes its units from inches to miles, and the forecaster simply carries out the appropriate inches to miles unit conversion, their point score will be unaffected. This is contrasted with the linear prediction interval scoring rule (as well as nearly all others), where we have the strange seeming property that the points given depend on the unit that the original question was phrased in.
 
 \end{itemize}
 
 Note that a potential drawback of the log prediction interval scoring rule is that it does not handle predictions for quantities that have a possibility of being zero or negative, since the formula depends on $\log(\frac{U}{L})$.
 \\
 \\
 As a rule of thumb, we think the linear prediction interval scoring rule is more appropriate for predictions that have similar scales to each other (e.g. a game where you have to predict what year in the 20th century an event occurred, or how many times different actors won an Oscar), whereas log prediction interval scoring is more appropriate when the predictions are guaranteed to be positive and span across many orders of magnitude, and where the main task is to get the rough order of magnitude correct (e.g. questions like "how many trees are there in the United States as of 2018?", or "How many babies are born in the world each day?").

 \subsection{Drawbacks}
 
However, both of the above prediction interval scoring rules have a number of drawbacks in the context of a practical calibration training system. Specifically:
 
 \begin{itemize}
 
 \item They will produce arbitrarily large point losses (i.e. outliers) when a forecasters prediction is bad enough. For instance, in the linear case there is no limit to how many points can be lost as $U-L$ grows and/or as $L-x$ grows. So a forecaster can completely destroy their previously good score with one arbitrarily bad prediction.
 
 \item In our testing they do not seem to produce point scores that have the intuitive relationships to each other when comparing predictions intervals of different percentiles (e.g. if a prediction game involves questions that use different values of $\beta$). This makes it hard to include prediction intervals of different percentiles in one training program and maintain a total point score that applies across different $\beta$'s.
 
 \item No matter how the free parameters d and c are set, receiving 0 points will not correspond to a prediction where the forecaster indicates a complete lack of information. Even worse, if the forecaster attempts to reveal their ignorance by predicting with an extremely wide prediction interval, they will receive an extremely large point penalty as $(U-L)$ or $\log(\frac{U}{L})$ grows very large!
 
\item No matter how the free parameters d and c are set, whether the forecaster receives positive or negative points cannot be solely determined by whether their prediction interval contains the true value. That means that forecasters will, for instance, sometimes receive positive points when their interval contains the true value, and other times receive negative points when their interval contains the true value, a circumstance that our testing suggests can be confusing to users.

\item When the true value is contained within the prediction interval, the same number of points is rewarded no matter how close the true answer is to the center of the prediction interval. We believe it is unintuitive to forecasters that having the true value fall literally at the boundary (i.e. $x=L$ or $x=U$) is not worse than having the true value fall directly in the middle of the prediction interval (i.e. $x=\frac{L+U}{2}$ in the linear case). By a common sense definition of being a good predictor, the latter seems clearly better than the former. Intuition suggests that when the true value falls in the prediction interval, you should receive more points the closer the true value is to the `middle' of the prediction interval.
 
 \end{itemize}

\section{Intuitive Scoring on Right/Wrong Probabilistic Questions}

In creating our calibration training program, we opted to include two fundamentally different `types' of predictions, which capture important elements of making probabilistic predictions, but without introducing large amounts of complexity into the user interface. 

\subsection{Choice Predictions}

For `Choice Predictions', we ask the user to indicate what they think the correct answer is to the question posed to them, and then ask them to assign a probability to their answer being correct (in the form of a percentage confidence). This works well for a variety of question formats, including:
 \begin{itemize}
 \item True or false. \textbf{Example}: did New York, USA have a larger metro population than Kolkata, India in 2016?
 \item 1 choice from a list of n options (where only one choice is deemed correct). Example: which of these philosophers wrote `the Republic'?
 \item k choices from a list of n options (where only one choice is deemed correct, and the probability you assign is for this correct option being among the k you chose). \textbf{Example}: What truth rating was given to the following statement made by Bernie Sanders on Monday, April 25th, 2016? (select two choices from the options - True, Mostly True, Half-True, Mostly False, False).
 \item Write a word or phrase (where each is deemed either correct or incorrect based on on a pre-determined rule, for instance by being a case insensitive match to a known answer). \textbf{Example}: What sport did the Philadelphia Atoms and the Dallas Tornado play in the 1970's?
 \item Write a number (where only one number is deemed correct). \textbf{Example}: How many stars are on the flag of New Zealand?
 \end{itemize}

An appealing aspect of Choice Predictions is that they don't require the user to describe the probability of each possible answer choice, only the choice that they actually selected. This substantially simplifies the user interface and reduces the burden and effort for the user. It also makes it possible to include open ended `write a word' and `write a number' question types, where describing the probability of every possible choice is infeasible.
\\
\\
A simple (but unfortunately problematic) scoring rule for `Choice Predictions' would be, for example, to use the log scoring rule:
\begin{equation*}
 \begin{cases} 
      -\log(p)   & , \hspace{3pt}correct\hspace{5pt}i.e.\hspace{5pt}c=1 \\
      -\log(1-p) & , \hspace{3pt}incorrect\hspace{5pt}i.e.\hspace{5pt}c=0
\end{cases}  
\end{equation*}
where $p$ is the (scalar, not vector) probability that the forecaster assigns to the choice they made being correct, and $c=1$ if they were correct in their choice, whereas $c=0$ if they were incorrect. While such a scoring rule has the advantage of being proper, it satisfies almost none of the other desirable properties that we would hope to have.

\subsection{Practical Scoring Rules}

Let $S(p, c)$ be a proper scoring rule for Choice Predictions. And lets assume that larger values correspond to performing better, so that the scoring function indicates how well the forecaster performed (which also means that $S(p, 1)$ is increasing in $p$, and that $S(p, 0)$ is decreasing in p). We can now transform this proper scoring rule to have more desirable properties from the point of view of creating a practical training system.
\\
\\
Given any proper scoring rule of the form $S(p, c)$, we define a corresponding `Practical' scoring rule:
\begin{equation*}
S^{*}(p, c) = S_{\max}  \frac{S(p, c) - S(p_{rand}, c)}{S(p_{\max}, 1) - S(p_{rand}, 1) }
\end{equation*} 
\begin{equation*}
= \frac{S_{\max}}{{S(p_{\max}, 1) - S(p_{rand}, 1) }}
\begin{cases} 
     S(p, 1) - S(p_{rand}, 1)   & , when\hspace{5pt}c=1 \\
     S(p, 0) - S(p_{rand}, 0) & , when\hspace{5pt}c=0
\end{cases}  
\end{equation*}
where $S_{max}$ is the maximum score ever awarded on a prediction (e.g. $S_{max} = 10$), $p_{max}$ is the maximum probability that you are allowed to give as an answer (which we assume is not necessarily 1), and $p_{rand}$ is whatever probability choose to correspond to a completely random guess (e.g. $p_{rand}=\frac{1}{n}$ when there are $n$ outcomes to choose from, equivalent to guessing at random between these options). Of course $p_{max} > p_{rand}$, and we assume the maximum number of points is positive (since maximizing negative points would be confusing to most forecasters). Hence, conveniently,
\begin{equation}
\frac{S_{\max}}{{S(p_{\max}, 1) - S(p_{rand}, 1) }} > 0.
\end{equation}
We also assume that the user interface either limits the user to only predicting probabilities at least $p_{rand}$ and at most $p_{max}$, or that any probability assigned below $p_{rand}$ is set to $p_{rand}$ automatically, and any probability assigned above $p_{max}$ is set to $p_{max}$ automatically.
\\
\\
In addition to having a variety of other desirable properties, the `Practical' scoring rule formula above produces a proper scoring rule as long as $S(p,c)$ is proper, even though it is not just a simple linear transformation of $S(p,c)$.
\\
\\
To see why $S^{*}(p,c)$ is proper, we note that from the point of view of the forecaster, the average score value achieved when assigning a probability $q$ to having selected the right choice, when they in fact believe the probability is $p$, is given by:
\begin{equation*}
E = p S^{*}(q, 1) + (1-p) S^{*}(1-q, 0)
\end{equation*}
\begin{equation*}
= \frac{S_{\max}}{{S(p_{\max}, 1) - S(p_{rand}, 1) }} (p (S(q, 1) - S(p_{rand}, 1)) + (1-p) (S(1-q, 0) - S(p_{rand}, 0)))
\end{equation*}
\begin{equation*}
= \frac{S_{\max}}{{S(p_{\max}, 1) - S(p_{rand}, 1) }} \left (\left ( p S(q, 1) +  (1-p) S(1-q, 0) \right ) - p S(p_{rand}, 1) - (1-p) S(p_{rand}, 0) \right ).
\end{equation*}
Now, considered as a function of $q$, the only portion of this formula that depends on $q$ is:
\begin{equation*}
p S(q, 1) +  (1-p) S(1-q, 0)
\end{equation*}
so the entire formula is maximized in $q$ precisely when this portion is maximized in $q$. However, by definition (since $S(p,c)$  is a proper scoring rule) this formula is maximized when $q=p$, and therefore, $S^{*}(p,c)$ is also a proper scoring rule! At least, this is true so long as the predictor never desires to assign a probability outside of the interval $[p_{rand}, p_{max}]$.
\\
\\
\\
In addition to being proper though, $S^{*}$ has a number of other desirable properties that make it practical for use in real calibration training:
\begin{itemize}
\item A higher score corresponds to making better predictions, which is natural when we think of the score as assigning points based on performance.

\item If your prediction expressed the maximum amount of uncertainty (i.e. guessing at random by assigning $p=p_{rand}$) then you get 0 points, which makes sense intuitively because this case is equivalent to abstaining from having an opinion.

\item The forecaster will only receive positive points when correct (since $S(p, 1) - S(p_{rand}, 1) $ is 0 when $p=p_{rand}$, and is increasing in p), and will only receive negative points when incorrect (since $S(p, 0) - S(p_{rand}, 0)$ is 0 when $p=p_{rand}$ and is decreasing in p). As mentioned, this is appealing because it helps forecasters quickly understand the outcome of a prediction (since the sign of the result makes it immediately obvious if they were correct or not).

\item By limiting our user interface to only allowing probabilities that are no greater than $p_{max}$, we prevent outliers in the formula and control the maximum number of points the forecaster can get in any one prediction to $S_{max}$, which occurs when they are correct (i.e. $c=1$) with the maximum allowed confidence ($p=p_{max}$).

\end{itemize}

\section{Our scoring rules}

\begin{figure}[h]
\caption{Practical score for binary probabilities $\{0.01, 0.02, \ldots, 1.00\}$} with $p_{max} = 1$ and $s_{max} = 2$
\includegraphics[width=0.7\textwidth]{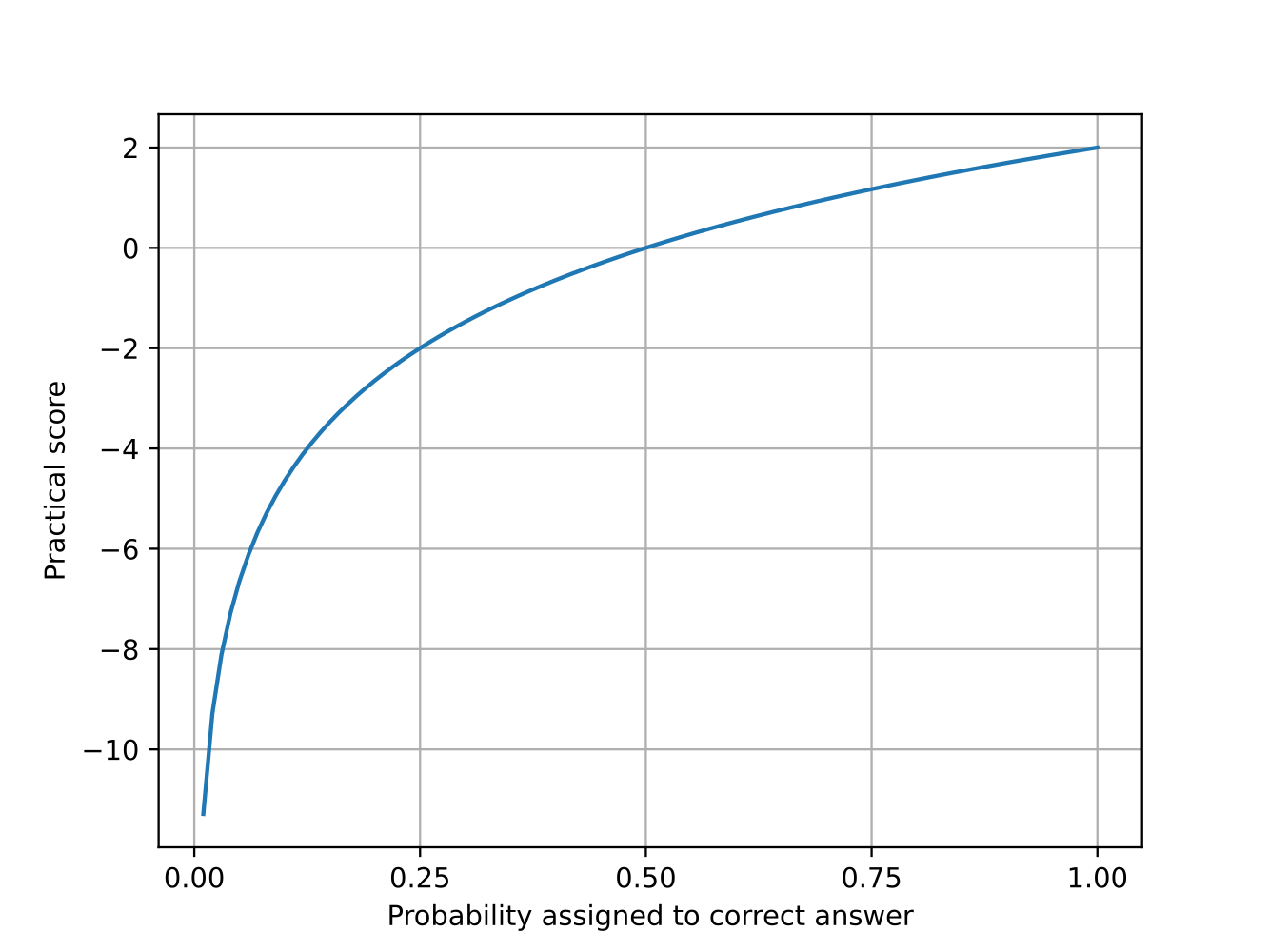}
\centering
\label{fig:practical}
\end{figure}

Due to the appealing properties of the logarithmic scoring rule mentioned previously, for the case of our calibration training program we chose to apply our Practical scoring rule formula to the log scoring rule, yielding:
\begin{equation*}
S(p, c) = \frac{s_{max}}{\log(p_{max}) - \log(p_{rand})} \hspace{2pt}*\hspace{2pt}
\begin{cases} 
      \log(p) -  \log(p_{rand})   & , \hspace{3pt}correct\hspace{5pt}i.e.\hspace{5pt}c=1 \\
      \log(1-p) -  \log(1-p_{rand})  & , \hspace{3pt}incorrect\hspace{5pt}i.e.\hspace{5pt}c=0
\end{cases}  
\end{equation*}
where $p_{rand}$ is the probability of getting the answer correct if you guess uniformly at random, $p_{max}$ is the maximum probability that the user is allowed to indicate for their confidence, and $s_{max}$ is the maximum number of points that a user can get on one question. Note that the base of the logarithm does not change the scoring function, because a change of base would effect the top and bottom equally, canceling out.
\\
\\
In the special case of a true/false question (or any question where there are only two answer options), we have that $p_{rand} = \frac{1}{2}$, so the scoring rule simplifies to:
\begin{equation*}
S(p) = \frac{s_{max}}{\log(p_{max}) - \log(p_{rand})} \hspace{2pt}*\hspace{2pt} \left ( 
\begin{cases} 
      \log(p)    & , \hspace{3pt}correct \\
      \log(1-p) & , \hspace{3pt}incorrect 
\end{cases} 
\hspace{5pt} -\hspace{3pt}  \log(1/2)
\right )  
\end{equation*}
This special case is just a simple linear transformation of the $log$ scoring rule, which is not true more generally. This is visualized in figure \ref{fig:practical}.

\subsection{Interval Predictions}

\subsubsection{Distance Scoring Rule}
When the order of magnitude of a prediction is obvious to the forecasters (such as is typically the case with guessing historical dates, or when predicting a percentage, etc.), 
it is reasonable to use a scoring rule that depends on the distance between the true value x, and the prediction interval (i.e. on $\frac{x-U}{c}$ or $\frac{L-x}{c}$), as well as the size of the prediction interval, given by $\frac{U-L}{c}$ (where $c$ determines the units with which we measure distance). Indeed, this is how the linear prediction interval scoring rule (equation \ref{eqn:linear_prediction_interval}) works, which recall is a proper scoring rule.  
\\
\\
Following this line of thinking (as well as for other reasons that we will elaborate on shortly), we define our Distance prediction interval scoring rule to be:

\begin{equation*}
S^0_{dist}(x, L, U) = 
\begin{cases} 
     \frac{-2}{1-\beta} \frac{L-x}{c}  - \frac{ \frac{L-x}{c} } {1 +  \frac{L-x}{c} } (\frac{U-L}{c}) & , when\hspace{5pt}x < L \\
     4 s_{max} (\frac{U-x}{U-L}) (\frac{x-L}{U-L}) (1 - \frac{\frac{U-L}{c}}{1+\frac{U-L}{c}})  &, when\hspace{5pt}L\leq x\leq U   \\
     \frac{-2}{1-\beta} \frac{x-U}{c}  - \frac{ \frac{x-U}{c} } {1 +  \frac{x-U}{c} } (\frac{U-L}{c}) & , when\hspace{5pt}x > U \\
\end{cases}  
\end{equation*}
Or, equivalently, we can write:
\begin{equation*}
S^0_{dist}(x, L, U) = 
\begin{cases} 
     \frac{-2}{1-\beta} r  - \frac{ r} {1 + r } s & , when\hspace{5pt}x < L \\
     4 s_{max} \frac{r \hspace{2pt} t}{s^2}  (1 - \frac{s}{1+s})  &, when\hspace{5pt}L\leq x\leq U   \\
     \frac{-2}{1-\beta} t  - \frac{ t } {1 + t } s & , when\hspace{5pt}x > U \\
\end{cases}  
\end{equation*}
where $r = \frac{L-x}{c}$, $s = \frac{U-L}{c}$, and $t = \frac{x-U}{c}$.

\subsubsection{Order of Magnitude Scoring Rule}
When the order of magnitude of a prediction is not obvious to the forecasters, then predicting this order of magnitude should be considered the main task of forecasting (such as is typical when predicting the number of people or things with a certain property, like the number of stars in the milky way galaxy, or the number of high school basketball teams in the United States). In such cases, it is reasonable to use a scoring rule that depends on how many orders of magnitude there are are between the true value x and the prediction interval (i.e. $\frac{\log(x/U)}{c}$ or $\frac{\log(L/x)}{c}$), as well as the number of orders of magnitude spanned by the prediction interval (i.e. $\frac{\log(U/L)}{c}$). Here, the constant $c$ determines the base of the logarithm we use to measure orders of magnitude (e.g. base 10 vs. base 2).
\\
\\
Following this line of thinking, we define our Order of Magnitude prediction interval scoring rule to be:

\begin{equation*}
S^0_{mag}(x, L, U) = 
\begin{cases} 
     \frac{-2}{1-\beta} \frac{\log(L/x)}{c}  - \frac{ \frac{\log(L/x)}{c} } {1 +  \frac{\log(L/x)}{c} } \frac{\log(U/L)}{c} & , when\hspace{5pt}x < L \\
     4 s_{max} \frac{\log(L/x)}{\log(U/L)} \frac{\log(x/U)}{\log(U/L)}  (1 - \frac{\frac{\log(U/L)}{c}}{1+\frac{\log(U/L)}{c}})  &, when\hspace{5pt}L\leq x\leq U   \\
     \frac{-2}{1-\beta} \frac{\log(x/U)}{c}  - \frac{ \frac{\log(x/U)}{c} } {1 +  \frac{\log(x/U)}{c} } \frac{\log(U/L)}{c} & , when\hspace{5pt}x > U \\
\end{cases}  
\end{equation*}

Or equivalently, we can write:
\begin{equation*}
S^0_{mag}(x, L, U) = 
\begin{cases} 
     \frac{-2}{1-\beta} r  - \frac{ r} {1 + r } s & , when\hspace{5pt}x < L \\
     4 s_{max} \frac{r \hspace{2pt} t}{s^2}  (1 - \frac{s}{1+s})  &, when\hspace{5pt}L\leq x\leq U   \\
     \frac{-2}{1-\beta} t  - \frac{ t } {1 + t } s & , when\hspace{5pt}x > U \\
\end{cases}  
\end{equation*}
where $r = \frac{\log(L/x)}{c}$, $s = \frac{\log(U/L)}{c}$, and $t = \frac{\log(x/U)}{c}$.
\\
\\
Note that this second formula is identical to the formula for our Distance scoring rule above, except for the definitions of $r$, $s$ and $t$, which have changed from linear relationships to logarithmic ratios.
\\
\\
Like the log prediction interval scoring rule, the Order of Magnitude scoring rule is unit invariant, so changing units (e.g. from inches to miles) will not change a forecaster's score, assuming they simply apply the unit conversion to their forecasts.

\subsection{Explanation}

We note that unfortunately both the Distance and Order of Magnitude scoring rules above are not proper scoring rules. This is a deficit of these formulas, but it is one taken on willingly in order to have the increased flexibility to achieve other desirable properties that the standard linear and log prediction interval scoring rules do not have. The Distance and Order of Magnitude scoring rules were designed to be similar in form to the linear and log prediction interval scoring rules, but targeting additional properties.
\\
\\
Consider this formula which represents both the Distance and Order of Magnitude scoring rules:
\begin{equation*}
S^0(x, L, U) =
\begin{cases} 
     \frac{-2}{1-\beta} r  - \frac{ r} {1 + r } s & , when\hspace{5pt}x < L \\
     4 s_{max} \frac{r \hspace{2pt} t}{s^2}  (1 - \frac{s}{1+s})  &, when\hspace{5pt}L\leq x\leq U   \\
     \frac{-2}{1-\beta} t  - \frac{ t } {1 + t } s & , when\hspace{5pt}x > U \\
\end{cases}.  
\end{equation*}
Compare this now to the analogous formula for the standard proper scoring rules for prediction intervals above (after we set the constant $d=0$ and renormalize the whole equation by multiplying by $\frac{2}{1-\beta}$):

\begin{equation}
S(x, L, U) =  
\begin{cases} 
     \frac{-2}{1-\beta} r - s & when\hspace{5pt} x<L \\
     -s & when\hspace{5pt} L \le x \le U \\
      \frac{-2}{1-\beta} t - s & when\hspace{5pt}x > U 
\end{cases} 
\end{equation}

The ways we modify the second formula to produce the first are by changing:
\begin{enumerate}
\item -s (in the case where $x < L$), which is replaced with $-\frac{r}{1+r} s$, so that we get 0 points when $x=L$ (i.e. when the true value is on the boundary of the prediction interval). Note that when $r$ is large it is the case that $\frac{r}{1+r} \approx 1 - \frac{1}{r}$, so for large $r$ (i.e. L being much bigger than the actual realized value) this modification has little effect on the formula.
\item -s (in the case where $x >U$), which is replaced with $-\frac{t}{1+t} s$, so that we get 0 points when $x=U$ (i.e. when the true value is on the boundary of the prediction interval). When $t$ is large (i.e. U is much too small) this modification has little effect on the formula.
\item -s (in the case where $L \le x \le U$), which is replaced with $4 s_{max} \frac{r \hspace{2pt} t}{s^2} (1-\frac{s}{1+s})$, so that (1) as the prediction interval size goes to infinity (i.e. $s \rightarrow \infty $) we get zero points (since an arbitrarily large prediction interval amounts to no guess at all), and so that (2) at the edge of the prediction interval (i.e. when $x=L$ or $x=U$) we get zero points (which also gives us continuity, since it causes the function to line up with the other pieces of the function at the boundaries), and so that (3) we get a maximum score of $s_{max}$, which occurs precisely when the true value lands right at the middle of the prediction interval (we consider the middle to be the arithmetic mean $x= \frac{L+U}{2}$ in the Distance case, and to be the geometric mean $x=\sqrt{L U}$ in the Order of Magnitude case). Note that the factor $4 \frac{r \hspace{2pt} t}{s^2}$ is specially chosen to satisfy $4 \frac{r \hspace{2pt} t}{s^2} = 0$ when either $x=L$ or $x=U$, with the maximum value of $4 \frac{r \hspace{2pt} t}{s^2}$ being $1$, occurring at the mean and geometric mean (in the Distance and Order of Magnitude cases, respectively).
\end{enumerate}
In other words, we use a scoring rule still heavily inspired by the proper scoring rule, even though it has lost the properness property. Fortunately, we have gained other desirable properties in the process that the linear and log prediction interval scoring rules lack, including:
\begin{enumerate}
\item a maximum (positive) score of $s_{max}$, which is achieved only in the case of the best possible forecast (i.e. in the limit as the prediction interval shrinks to zero size, while centered perfectly at the true value).
\item zero points when the true value falls at either prediction interval boundary (i.e. when $x=L$ or $x=U$).
\item a limit of zero points as the interval size approaches infinity (i.e. a completely non-informative prediction).
\item a continuous function (i.e. no `jump' in score for arbitrarily small changes in the prediction interval boundaries $L$ and $U$).
\item a score that gets bigger the closer the true prediction is to the center of the prediction interval (unlike in the proper scoring rule cases where the same score is always achieved, for a given prediction interval size, whenever the true value falls in the prediction interval).
\end{enumerate}

\subsection{Final prediction interval scoring rules}

Both the `Distance' and `Order of Magnitude' scoring rules above still suffer from the following drawbacks:
 \begin{enumerate}
 \item If the true value lies right on the edge of the prediction interval, it gives strictly zero points. In our experience this can confuse users because they feel such a prediction should still be considered `correct', and hence, should give positive points.
 \item There is no limit to the most points that can be lost (i.e. negative outliers are uncapped).
 \end{enumerate}

To resolve both of the above issues, we introduce our final prediction interval scoring rules, which are built from these two prior rules by making slight modifications.

\begin{equation*}
S_{dist}(x,L,U) =
\begin{cases} 
S^0_{dist}(x, L-\delta, U+\delta) & , when\hspace{5pt}S^0_{dist}(x, L-\delta, U+\delta) > s_{min} \\
s_{min} &, otherwise
\end{cases}
\end{equation*}

\begin{equation*}
S_{mag}(x,L,U) =
\begin{cases} 
S^0_{mag}(x, L (1-\delta), U (1+\delta)) & , when\hspace{5pt}S^0_{mag}(x, L (1-\delta), U (1+\delta)) > s_{min} \\
s_{min} &, otherwise
\end{cases}
\end{equation*}

In words, these final scoring rules simply set the score to $s_{min}$ if the score would otherwise be below $s_{min}$, and also slightly inflate the prediction interval size (by a small `expansion factor' $\delta$), so that if the true value falls right at the edge of the original prediction interval, the forecaster still gets positive points.

\subsection{Parameter Choices}

The the 'Practical', `Distance' and `Order of Magnitude' scoring rules, which we preferred for use in our calibration training program, have a number of parameters that need to be set. We considered various options and settled on:
\begin{enumerate}
\item The maximum number of points per forecast was set as $s_{max} = 10$. We think it is intuitive and natural for users to understand the idea that the most points they can get on any single forecast is 10.
\item The maximum probability the forecaster is allowed to use in a prediction was set as $p_{max} = 0.99$, and correspondingly, the minimum probability as $1-p_{max} = 0.01$. Allowing very high confidence and very low confidence predictions is problematic because (a) it will produce outliers with log based scoring rules, (b) in practice it is very difficult for people to intuitively grasp differences in probabilities that are very small in absolute value, and (c) it would take an unrealistically large number of rounds of training practice to train yourself to be calibrated on events that occur, say, only 1 in 1000 times. Hence, we simply do not give the user the option of selecting extremely high confidences or extremely low confidences in our user interface.
\item We set the most points you can lose per forecast as $s_{min}$ $= -57.26893683880667$ \\
$= -((10 \log(99/50))/\log(50))$. This (perhaps seemingly odd) choice is the natural maximum point loss when using our practical scoring rule with the selected parameter values $p_{max} = 0.99$ and $s_{max} = 10$ on true/false prediction problems. We chose to maintain consistency across prediction types by using this value for $s_{min}$ in every case that it applied.
\item The prediction interval `expansion factor' was set as $\delta=0.4$. For integer predictions this will not significantly change your score, but it does cause users to get positive points in the case where their predictions fall right on the prediction interval boundary, which in user testing seemed to be preferable to giving strictly zero points in that case (e.g. if the user gives the prediction interval [10,100], and the actual value is 10, the user expects to receive at least some positive reward for this).
\item The scale parameter was set as $c=100$ (for questions with `Distance' scoring) and $c=ln(100)=4.60517$ (for questions with `Order of Magnitude' scoring). This parameter determines for prediction interval forecasts what is to be considered a `large' versus `small' mistake. In this instance, we consider a mis-prediction by 100 (or, an over or under prediction of two orders of magnitude) a moderately large mistake.
\end{enumerate}

\section{Conclusion}
\label{sec:conclus}

Proper scoring rules have the appealing property that they incentivize honest reporting of probabilistic beliefs. But in the context of creating a practical calibration training system, where real people train on prediction problems, there are other additional properties that we might desire a scoring rule to satisfy. We enumerated various properties that might be desired in this context, and introduced a subset of proper scoring rules, called `Practical' scoring rules, which in the context of grading a probabilistic `right' or `wrong' prediction has many of these proposed properties. Moreover, in the context of forecasters predicting numerical values by providing prediction intervals, we introduced the `Distance' and `Order of Magnitude' scoring rules. While these latter rules unfortunately sacrifice properness, they correct a number of deficiencies of the standard linear and log prediction interval scoring rules that are potentially problematic in a realistic training context where user experience is important.

\bibliography{calibration}

\end{document}